Superconductivity of $M_I(M_{II0.5}, Si_{0.5})_2$ ($M_I$=Sr and Ba, $M_{II}$=Al and Ga), ternary silicides with the $AlB_2$-type structure

Motoharu Imai, Kenji Nishida, Takashi Kimura, Hideaki Kitazawa, Hideki Abe
National Institute for Materials Science
1-2-1 Sengen, Tsukuba, Ibaraki 305-0047, JAPAN
Hijiri Kito
National Institute of Advanced Industrial Science and Technology
1-1-1 Umezono, Tsukuba, Ibaraki 305-8568, JAPAN
Kenji Yoshii
Japan Atomic Energy Research Institute
Mikazuki, Hyogo 679-5148 JAPAN

Ternary silicides $M_I(M_{II0.5}, Si_{0.5})_2$ ($M_I$=Sr and Ba, $M_{II}$=Al and Ga) were prepared by Ar arc melting. Powder X-ray diffraction indicates that they have the $AlB_2$-type structure, in which Si and $M_{II}$ atoms are arranged in honeycomb layers and $M_I$ atoms are intercalated between them. Electrical resistivity and dc magnetization measurements revealed that $Sr(Al_{0.5}, Si_{0.5})_2$ is superconductive, with a critical temperature for superconductivity ($T_C$) of 4.2 K, while $Ba(Al_{0.5}, Si_{0.5})_2$ is not at temperatures ranging above 2.0 K. $Sr(Ga_{0.5}, Si_{0.5})_2$ and $Ba(Ga_{0.5}, Si_{0.5})_2$ are also superconductors, with $T_C$s of 5.1 and 3.3 K, respectively.

Corresponding Author:
Motoharu IMAI
National Institute for Materials Science
1-2-1 Sengen,
Tsukuba, IBARAKI 305-0047, JAPAN
E-mail:IMAI.Motoharu@nims.go.jp
Tel: +81-298-59-2814, Fax: +81-298-59-2801

## 1. Introduction

Superconductivity at 39 K was recently reported with an intermetallic compound, $MgB_2$ [1]. Magnesium diboride has the $AlB_2$-type (C32-type) structure and, therefore, related compounds are attracting attention because of their potential as new superconductors. The discovery and study of new superconductors with the $AlB_2$-type structure are relevant to our understanding of the origin of the superconductivity of $MgB_2$. Silicides, as well as borides, also contain compounds with the $AlB_2$-type structure or a structure related to it.

Of binary silicides, $ThSi_2$ [2], $USi_2$ [3], and several rare-earth metal disilicides [4, 5] are known to have an $AlB_2$-type structure. Among them, $\beta$-$ThSi_2$ is known to be a superconductor with a critical temperature $T_C$ of 2.41 K [6]. A high-pressure phase of $CaSi_2$, which appears above 16 GPa, has an $AlB_2$-like structure in which Si atoms form slightly corrugated honeycomb layers [7]. This phase of $CaSi_2$ is superconductive, with a $T_C$ of 14 K at such pressures [8], while the ambient phase is not [9]. This high-pressure phase cannot be quenched at ambient conditions.

Recently, ternary silicides $M_{AE}(Ga_x, Si_{1-x})_2$ ($M_{AE}$=Ca, Sr, and Ba) have been reported to have the $AlB_2$-type structure and to be superconductors with a $T_C$ ranging from 3.3 to 3.9 K [10, 11]. Furthermore, a ternary silicide, $Ca(Al_{0.5}, Si_{0.5})_2$, which has the $AlB_2$-type structure, has been reported to be a superconductor with a $T_C$ of 7.7 K [12]. Ternary silicides $Sr(Al_x, Si_{1-x})_2$ and $Ba(Al_x, Si_{1-x})_2$ have been reported to have the $AlB_2$-type structure [13-16], though their electrical properties have not been examined yet. It would be interesting to examine whether the $Sr(Al_x, Si_{1-x})_2$ and $Ba(Al_x, Si_{1-x})_2$ show superconductivity, as $M_{AE}(Ga_x, Si_{1-x})_2$ and $Ca(Al_{0.5}, Si_{0.5})_2$ do.

Here, we report the synthesis of ternary silicides $Sr(Al_{0.5}, Si_{0.5})_2$ and $Ba(Al_{0.5}, Si_{0.5})_2$ and demonstrate that $Sr(Al_{0.5}, Si_{0.5})_2$ is a superconductor. We also investigated $Sr(Ga_{0.5}, Si_{0.5})_2$ and $Ba(Ga_{0.5}, Si_{0.5})_2$ for comparison.

## 2. Experimental

The samples were prepared by Ar arc melting of 1 : 1 : 1 stoichiometric mixtures of $M_I$ ($M_I$= Sr and Ba, nominal purity 99.5 %), $M_{II}$ ($M_{II}$=Al and Ga, nominal purity 99.999 %), and Si (nominal purity 99.9999 %). In the following, ARC-$M_I M_{II}$ denotes the sample prepared by Ar arc melting of 1 : 1 : 1 stoichiometric mixtures of $M_I$, $M_{II}$, and Si. Their chemical compositions were determined using electron probe microanalysis (EPMA). The samples for EPMA were prepared by polishing with an oil-based diamond slurry to avoid a reaction with water. The structure of the silicides was examined using a powder X-ray diffraction method. Electrical resistivity was measured by means of the four-probe method at temperatures ranging from 2 to 290 K. Since ARC-SrAl and ARC-BaAl were slightly sensitive to moisture in air, they were covered with insulating varnish (General Electric, 7031) after spark-bonding of Au wires. dc magnetization was measured using a SQUID magnetometer at temperatures ranging from 2 to 12 K.

## 3. Results

Map analysis by EPMA shows that all the samples consist of almost a single phase, though they contain small amounts of precipitates. Table I lists the average chemical composition of the main phase determined with EPMA, which confirms that the $M_I(M_{II0.5}Si_{0.5})_2$ phase is the main one in all the samples. A small amount of oxygen was detected in ARC-BaGa. This would be, however, due to the surface oxidization of samples because the amount of oxygen detected was small and it was also detected in the precipitates.

Figure 1 shows the powder X-ray diffraction patterns of the samples and calculated diffraction patterns for the $M_I(M_{II0.5}Si_{0.5})_2$ phase with the $AlB_2$-type structure. For the calculation, we assumed that the compound $M_I(M_{II0.5}, Si_{0.5})_2$ would have the $AlB_2$-type structure with the observed lattice constants described below, in which the 2d site is chemically disordered with occupation probabilities of 0.5 for the element $M_{II}$ and 0.5 for Si. The major observed reflections are reproduced well by the assumed structure, as shown in Fig. 1. These results indicate that the $M_I(M_{II0.5}Si_{0.5})_2$ phase has the $AlB_2$-type structure.

Their lattice constants are listed in Table II, together with those of $Ca(Al_{0.5}, Si_{0.5})_2$ [12] and $Ca(Ga_{0.5}, Si_{0.5})_2$ [11]. When the alkaline-earth elements are changed, the lattice constants change anisotropically: the lattice constant $a$ increases by 2.5 % for $M_I(Al_{0.5}, Si_{0.5})_2$ and by 3.3 % for $M_I(Ga_{0.5}, Si_{0.5})_2$, while the lattice constant $c$ increases by 16.9 % for $M_I(Al_{0.5}, Si_{0.5})_2$ and 14.9 % for $M_I(Ga_{0.5}, Si_{0.5})_2$ when the alkaline-earth elements are changed in the sequence Ca, Sr, and Ba.

Figures 2 and 3 show the electrical resistivity of $M_I(Al_{0.5}, Si_{0.5})_2$ and $M_I(Ga_{0.5}, Si_{0.5})_2$ as a function of the temperature, respectively. The resistivity starts to decrease at temperature $T_C^{ONSET}$ and becomes negligibly small at temperature $T_C^{ZERO}$ in all the samples except for $Ba(Al_{0.5}, Si_{0.5})_2$. Table III lists $T_C^{ONSET}$ and $T_C^{ZERO}$ together with those of related materials. The resistivity of $Ba(Al_{0.5}, Si_{0.5})_2$ decreases continuously with decreasing the temperature down to 2 K, the lowest temperature of this experiment. These results suggest that the silicides $M_I(M_{II0.5}, Si_{0.5})_2$ become superconductive at $T_C^{ZERO}$ except for $Ba(Al_{0.5}, Si_{0.5})_2$.

Figure 4 shows the dc magnetization $M$ of $M_I(M_{II0.5}, Si_{0.5})_2$ except for $Ba(Al_{0.5}, Si_{0.5})_2$ in a field $H$ as a function of the temperature. The triangles and circles show data obtained in zero-field cooling (ZFC) and those in field cooling (FC), respectively. The samples show a Meissner effect (flux exclusion) below a temperature $T_C^{MEIS}$ in FC, which confirms the presence of a superconducting phase below the temperature, as indicated by the resistivity measurements. Table III also lists $T_C^{MEIS}$, the magnetic shielding fraction in ZFC, and the flux exclusion in FC at 2.0 K. These indicate that the superconductivity is a bulk effect. On this basis, we conclude that ternary silicides $M_I(M_{II0.5}, Si_{0.5})_2$ are superconductors with the $AlB_2$-type structure, except for $Ba(Al_{0.5}, Si_{0.5})_2$. Therefore, it is revealed that six ternary $M_{AE}$-$M_{II}$-Si systems ($M_{AE}$=Ca, Sr, and Ba, $M_{II}$=Al and Ga) have compounds with the $AlB_2$-type structure, and the compounds in the systems are superconductive, except for that in the Ba-Al-Si system.

## 4. Discussion

In the silicides with the same third element $M_{II}$, the $T_C$ depends on the choice of $M_{AE}$. In the silicides $M_{AE}(Al_{0.5}, Si_{0.5})_2$, the $T_C$ decreases when the element $M_{AE}$ is changed from Ca into Sr and finally becomes lower than 2 K, the lowest temperature in this study, when the element $M_{AE}$ is Ba. On the other hand, in the silicides $M_{AE}(Ga_{0.5}, Si_{0.5})_2$, the $T_C$ becomes the highest when the element $M_{AE}$ is Sr.

In the Bardeen-Cooper-Schrieffer theory, the $T_C$ is estimated as

$$T_C = 1.13\Theta_D \exp(-1/N(E_F)V),$$

where $\Theta_D$ is the Debye frequency, $N(E_F)$ is the density of states at the Fermi level, and $V$ is the average electron pairing interaction [17]. In the silicides $M_{AE}(B_{0.5}, Si_{0.5})_2$, the Debye temperature is expected to be the same because the composition of honeycomb layers is the same in the three compounds. Therefore, the different $T_C$ in $M_{AE}(M_{II0.5}, Si_{0.5})_2$ would be attributed to the change in $N(E_F)$ under the assumption that $V$ is independent of the choice of $M_{AE}$. The change of the element $M_{AE}$ affects $N(E_F)$ in two ways. One is the effect on the Fermi energy. When the element $M_{AE}$ is changed in the sequence Ca, Sr, and Ba, the volume of the unit cell increases, as shown in Table II, and the valence electron density decreases. Consequently, the Fermi energy decreases in the sequence because the Fermi energy is proportional to 2/3 the power of the valence electron density. For instance, the difference in the Fermi energy between $Ca(Al_{0.5}, Si_{0.5})_2$ and $Ba(Al_{0.5}, Si_{0.5})_2$ is roughly estimated to be 13 %. The other is an effect on the shape of the density of states (DOS). By analogy with the calculated electronic structure of $CaSi_2$ with the $AlB_2$-type structure [18, 19], the $N(E_F)$ of $M_{AE}(M_{II0.5}, Si_{0.5})_2$ consists of mixed states of $d$ states of the element $M_{AE}$ and $p$ states of the element $M_{II}$ and Si. The $d$ states are expected to shift upwards when the element $M_{AE}$ is changed in the sequence Ca, Sr, and Ba. The upward shift of the $d$ states changes the shape of DOS. Therefore, the change of the element $M_{AE}$ causes the shift of the Fermi energy, the change of the DOS shape, and, consequently, the change of $N(E_F)$. This would be reflected on the difference in the $T_C$ among $M_{AE}(M_{II\ 0.5}, Si_{0.5})_2$ with the same third element $M_{II}$.

In summary, the ternary silicides $M_I(M_{II0.5}, Si_{0.5})_2$ were found to be superconductors with the $AlB_2$-type structure, except for $Ba(Al_{0.5}, Si_{0.5})_2$. Therefore, silicides with the $AlB_2$-type structure in $M_{AE}$-$M_{II}$-Si systems are concluded to be superconductive, except for that in the Ba-Al-Si system. On the other hand, in the binary alkaline-earth metal-boron system, only Mg forms a diboride with the $AlB_2$-type structure. In this sence, these ternary systems are in contrast with the binary alkaline-earth metal-boron system: the ternary alkaline-earth silicides have wider variations of the compounds with the $AlB_2$-typestructure than the binary alkaline-earth metal diborides.

**Acknowledgments**

We thank A. P. Tsai of NIMS for allowing us to use their arc furnace, and I. Nakatani and H. Mamiya of NIMS for allowing us to use the Physical Properties Measurement System (Quantum Design).

**References**
[1] J. Nagamatsu, N. Nakagawa, T. Muranaka, Y. Zenitani, and J. Akimitsu, Nature 410 (2001), 63.
[2] W. H. Zachariasen, Acta Cryst. 2 (1949), 94.
[3] A. Brown and J. J. Norreys, Nature 183 (1959), 673.

[4] I. P. Mayer, E. Banks, and B. Post, J. Phys. Chem. 66 (1962), 693.
[5] A. G. Tharp, J. Phys. Chem. 66 (1962), 758.
[6] G. F. Hardy and J. K. Hulm, Phys. Rev. 93 (1954), 1004.
[7] P. Bordet, M. Affronte, S. Sanfilippo, M. Núñez-Regueiro, O. Laborde, G. L. Olcese, A. Palenzona, S. LeFloch, D. Levy, and M. Hanfland, Phys. Rev. B 62 (2000), 11392.
[8] S. Sanfilippo, H. Elsinger, M. Núñez-Regueiro, O. Laborde, S. LeFloch, M. Affronte, G. L. Olcese, and A. Palenzona, Phys. Rev. B 61 (2000), R3800.
[9] D. B. MacWhan, V. B. Compton, M. S. Silverman, and J. R. Soulen, J. Less-Common Met. 12 (1967), 75.
[10] M. Imai, E. Abe, J. Ye, K. Nishida, T. Kimura, K. Honma, H. Abe, and H. Kitazawa, Phys. Rev. Lett. 87 (2001), 077003.
[11] M. Imai, K. Nishida, T. Kimura, and H. Abe, Physica C (in press).
[12] M. Imai, K. Nishida, T. Kimura, and H. Abe, Appl. Phys. Lett. 80 (2002), 1019.
[13] O. I. Bodak, E. I. Gladyshevskii, O. S. Zarechnyuk, and E. E. Cherkashin, Visn. Lviv. Gos. Univ. Ser. Khim. 8 (1965), 75 (in Russian) ; I. N. Ganiyev, A. V. Vakhobov, and T. D. Dzhurayev, Dolk. Akad. Nauk. Tazh. SSR 18 (1975), 27 (in Russian); G. Petzow and G. Effenberg, Ternary Alloys (VHC, Weinheim, 1990) Vol. 8.
[14] I. N. Ganiyev, A. V. Vakhobov, and T. D. Dzhurayev, Russian Metallurgy 4 (1977), 175.
[15] M. L. Fornasini and G. Bruzzone, J. Less-Common Met. 40 (1975), 335.
[16] I. N. Ganiyev, A. V. Vakhobov, and T. D. Dzhurayev, Russian Metallurgy 4 (1978), 183; G. Petzow and G. Effenberg, Ternary Alloys (VHC, Weinheim, 1990) Vol. 3.
[17] For example, M. Tinkham, *Introduction to Superconductivity*, 2nd ed., (McGraw-Hill, New York, 1996).
[18] K. Kusakabe, Y. Tateyama, T. Ogitsu, and S. Tsuneyuki, Rev. High Pressure Sci. Technol. 7 (1998), 192.
[19] G. Satta, G. Profeta, F. Bernardini, A. Continenza, and S. Massidda, Phys. Rev. B64 (2001), 104507.

Table I. Chemical compositions of the main phase in the samples.

| Sample | $M_I$ (at. %) | $M_{II}$ (at. %) | Si (at. %) | Phase |
|---|---|---|---|---|
| ARC-SrAl | 31.8 | 33.0 | 35.2 | $Sr(Al_{0.5}Si_{0.5})_2$ |
| ARC-SrGa | 33.8 | 33.2 | 33.0 | $Sr(Ga_{0.5}Si_{0.5})_2$ |
| ARC-BaAl | 32.7 | 33.3 | 34.0 | $Ba(Al_{0.5}Si_{0.5})_2$ |
| ARC-BaGa[a] | 33.4 | 33.8 | 30.9 | $Ba(Ga_{0.5}Si_{0.5})_2$ |

a) 1.8 at.% oxygen was detected in ARC-BaGa.

Table II. Lattice constants and theoretical density of $M_I(M_{II0.5}Si_{0.5})_2$ phases, together with those of $Ca(Al_{0.5}, Si_{0.5})_2$ and $Ca(Ga_{0.5}, Si_{0.5})_2$ phases.

| Phase | a(Å) | c(Å) | Volume (Å$^3$) | Theoretical density (gcm$^{-3}$) |
|---|---|---|---|---|
| $Ca(Al_{0.5}Si_{0.5})_2$[a] | 4.1905(5) | 4.3992(8) | 66.90 | 2.37 |
| $Sr(Al_{0.5}Si_{0.5})_2$ | 4.2475(3) | 4.7421(6) | 74.09 | 3.14 |
| $Ba(Al_{0.5}Si_{0.5})_2$ | 4.2973(8) | 5.1424(17) | 82.24 | 3.84 |
| $Ca(Ga_{0.5}Si_{0.5})_2$[b] | 4.1200(8) | 4.4401(1) | 65.27 | 3.43 |
| $Sr(Ga_{0.5}Si_{0.5})_2$ | 4.1875(4) | 4.7447(4) | 72.05 | 4.29 |
| $Ba(Ga_{0.5}Si_{0.5})_2$ | 4.2587(4) | 5.1039(9) | 80.17 | 4.85 |

a) Reference 12.
b) Reference 11.

Table III. Parameters of silicides $M_{AE}(M_{IIx}, Si_{1-x})_2$ ($M_{AE}$=Ca, Sr, and Ba, $M_{II}$=Al and Ga) observed in electrical resistivity and magnetization measurements. Units used in the magnetic shielding fraction and flux exclusion are a percentage of the theoretical value of the perfect diamagnetism ($1/4\pi$).

| Silicide | $T_C^{ONSET}$ (K) | $T_C^{ZERO}$ (K) | $T_C^{MEIS}$ (K) | Magnetic shielding fraction in ZFC at 2 K (%) | Flux exclusion in FC at 2 K (%) |
|---|---|---|---|---|---|
| $Ca(Al_{0.5}, Si_{0.5})_2$ [a] | 8.0 | 7.7 | 7.7 | 60 | 7 |
| $Sr(Al_{0.5}, Si_{0.5})_2$ | 4.3 | 4.2 | 4.2 | >90 | 8 |
| $Ca(Ga_{0.5}, Si_{0.5})_2$ [b] | 4.5 | 3.3 | 4.3 | 90 | 3 |
| $Ca(Ga_{0.37}, Si_{0.63})_2$ [b] | 3.6 | 3.5 | 3.5 | 80 | 0.6 |
| $Sr(Ga_{0.5}, Si_{0.5})_2$ | 5.2 | 5.1 | 5.1 | 90 | 8 |
| $Sr(Ga_{0.37}, Si_{0.63})_2$ [b] | 4.7 | 3.6 | 4.2 | 90 | 4 |
| $Sr(Ga_{0.37}, Si_{0.63})_2$ [c] | 3.5 | 3.4 | 3.4 | 80 | 8 |
| $Ba(Ga_{0.50}, Si_{0.50})_2$ | 4.1 | 3.3 | 3.9 | 90 | 8 |
| $Ba(Ga_{0.39}, Si_{0.61})_2$ [b] | 4.3 | 3.9 | 4.0 | >90 | 6 |

a) Reference 12.
b) Reference 11.
c) Reference 10.

Fig. 1. Powder X-ray diffraction patterns of $M_I(M_{II0.5}, Si_{0.5})_2$ ($M_I$=Sr and Ba, $M_{II}$=Al and Ga).

Fig. 2. Electrical resistivity of $M_I(Al_{0.5}, Si_{0.5})_2$ ($M_I$=Sr and Ba) as a function of the temperature.

Fig. 3. Electrical resistivity of $M_I(Ga_{0.5}, Si_{0.5})_2$ ($M_I$=Sr and Ba) as a function of the temperature.

Fig. 4. dc magnetization $M$ of $Sr(Al_{0.5}, Si_{0.5})_2$, $Sr(Ga_{0.5}, Si_{0.5})_2$, and $Ba(Ga_{0.5}, Si_{0.5})_2$ measured in field $H$ as a function of the temperature. The abbreviations "ZFC" and "FC" represent the magnetization measured in zero-field cooling and field cooling, respectively.

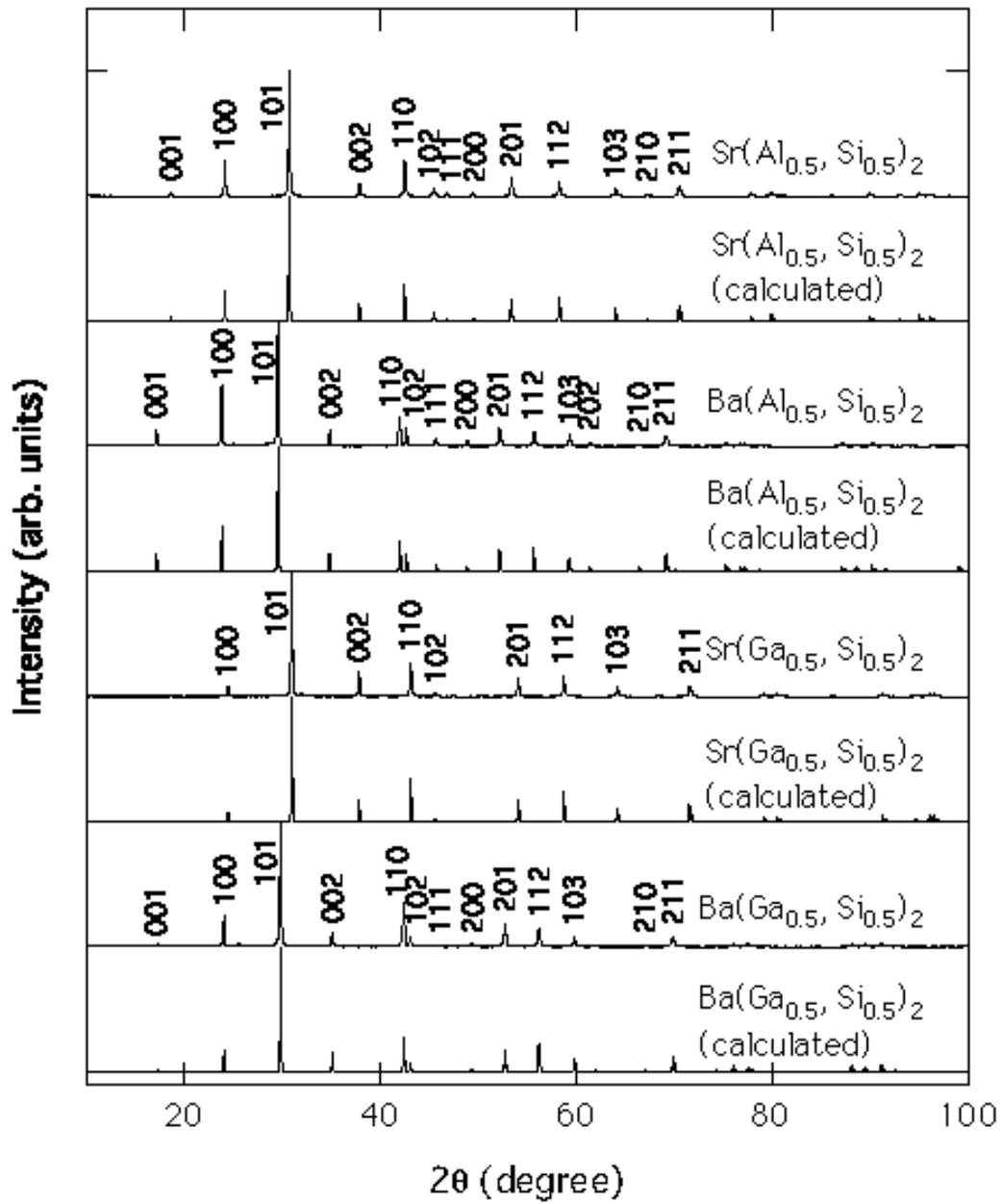

Fig.1 M. Imai et al.

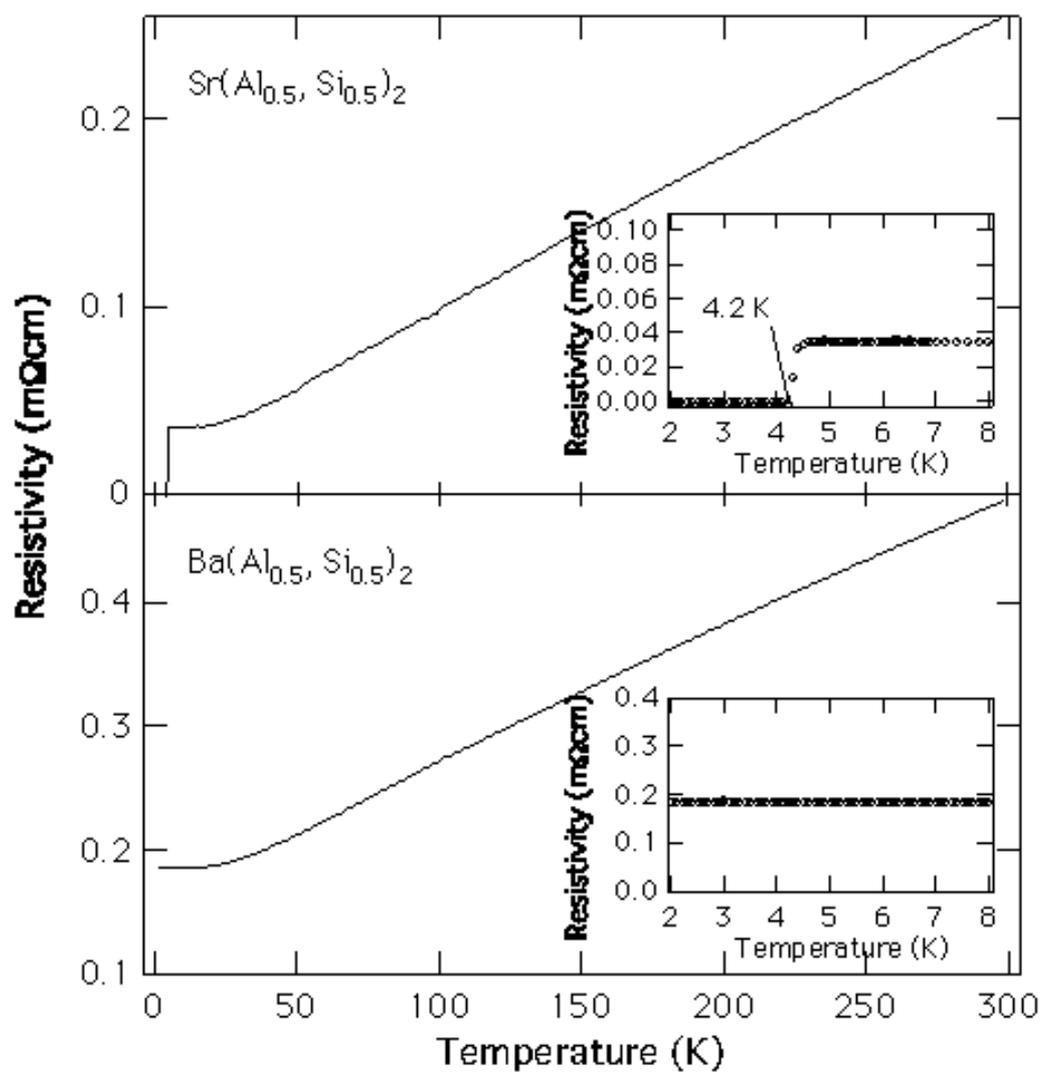

Fig.2 M. Imai et al.

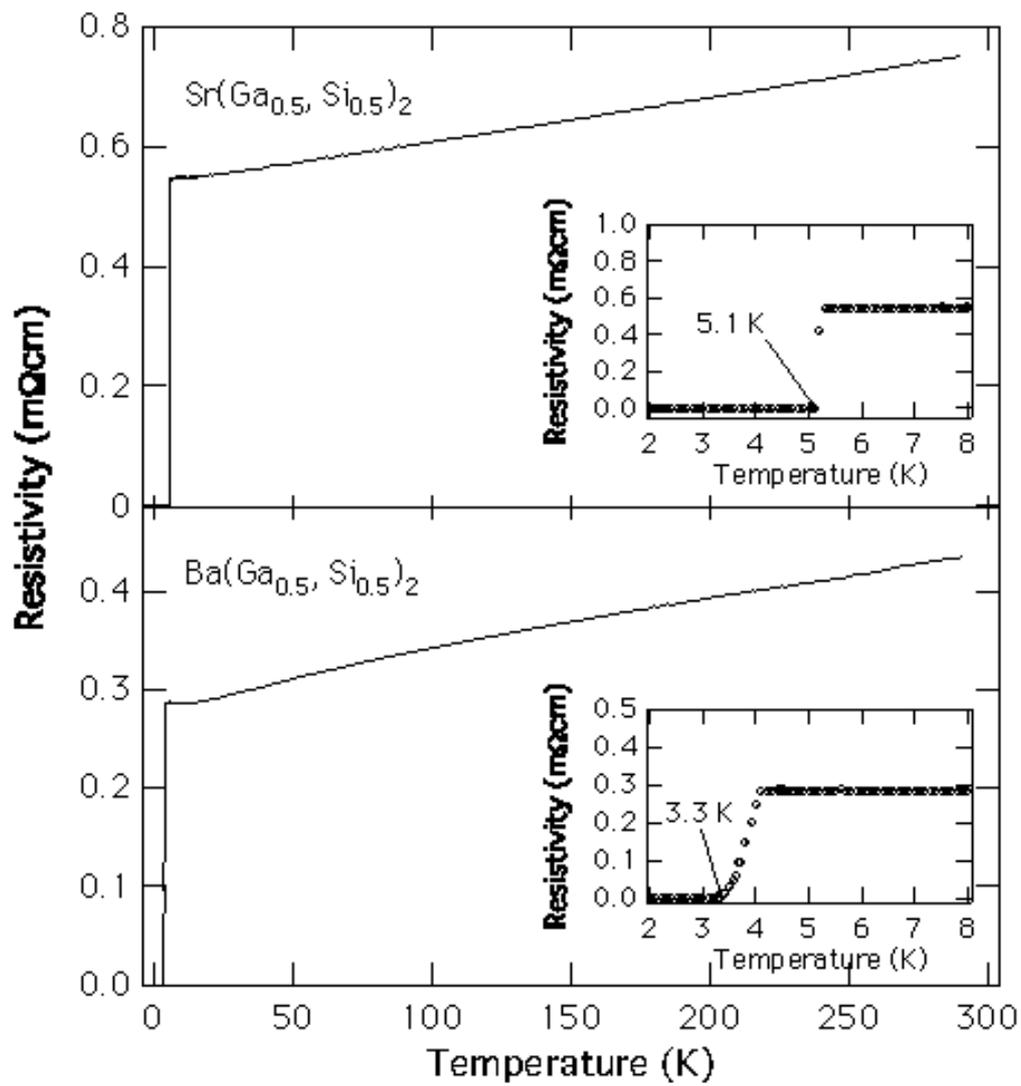

Fig.3 M. Imai et al.

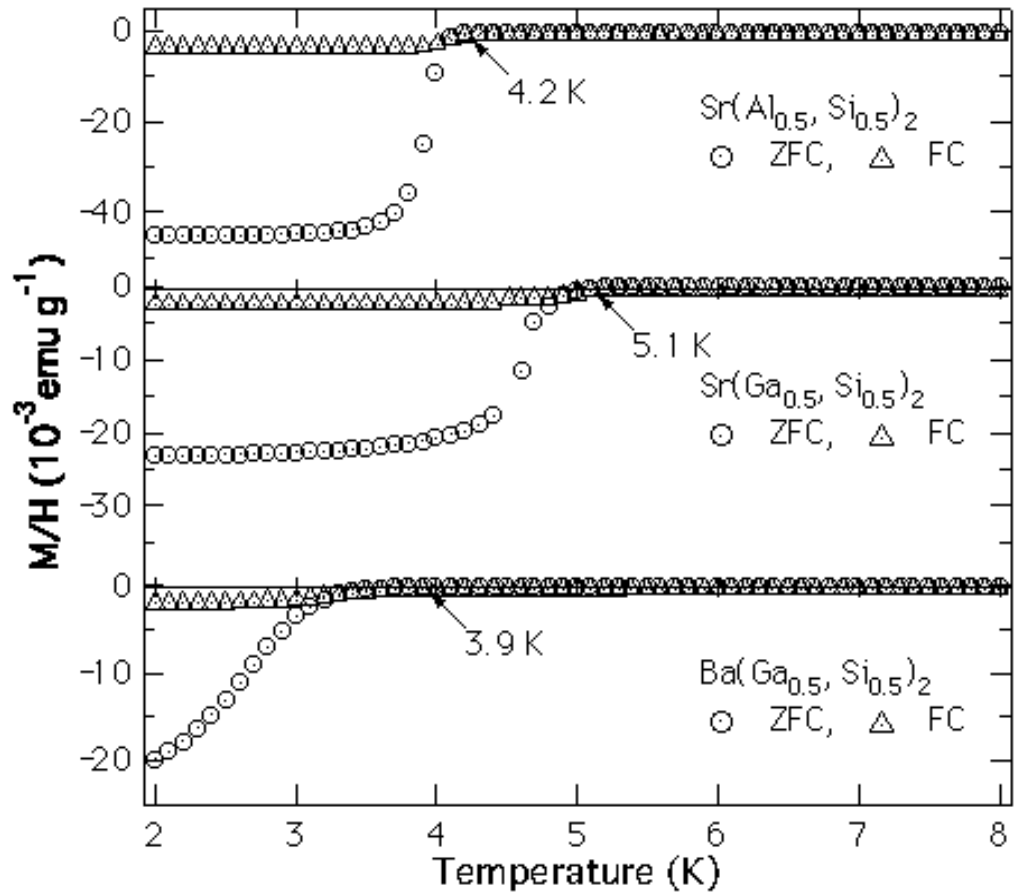

Fig. 4 M. Imai et al.